\documentclass[conference]{IEEEtran}
\IEEEoverridecommandlockouts
\usepackage{cite}
\usepackage{amsmath,amssymb,amsfonts}
\usepackage{algorithmic}
\usepackage{graphicx}
\usepackage{textcomp}
\usepackage{xcolor}
\def\BibTeX{{\rm B\kern-.05em{\sc i\kern-.025em b}\kern-.08em
    T\kern-.1667em\lower.7ex\hbox{E}\kern-.125emX}}
\begin{document}

\title{A Multimodal Machine Learning Framework for Enterprise Database Workload-Aware Root Cause Analysis}

\author{
\IEEEauthorblockN{1\textsuperscript{st} Ruchi Pakhle}
\IEEEauthorblockA{
\textit{Data and AI}\\
\textit{Red Hat}\\
rpakhle@redhat.com
}
\and
\IEEEauthorblockN{2\textsuperscript{nd} Siddhant Pawar}
\IEEEauthorblockA{
\textit{IBM}\\
sipawar@ibm.com
}
}

\maketitle

\begin{abstract}
The difficult and time-consuming process of root cause analysis (RCA) for corporate database events frequently depends on human examination of logs, performance indicators, and query activity. Because they usually concentrate on a single data source, like logs or system metrics, traditional incident diagnosis techniques are unable to fully capture the operational context of database issues. In reality, workload characteristics and observable system signals have a significant impact on events like lock contention, deadlocks, sluggish queries, CPU saturation, and I/O bottlenecks.

Three complimentary forms of evidence are used in the suggested method: workload patterns determined from SQL query behavior, time-varying performance measures, and log-derived representations. Unsupervised learning is used to first classify workloads into behavioral clusters. These workload-aware features are then combined with database metrics and logs to anticipate root causes. In addition to offering operational insight into how different workload types contribute to incident generation, the framework is intended to increase diagnostic accuracy. 

It is anticipated that the findings will show that taking workload behavior into account results in more precise and operationally significant incident diagnosis.
\end{abstract}


\section{Introduction}

In order to reduce downtime and service degradation, common problems including lock contention, deadlocks, sluggish query execution, CPU saturation, and I/O bottlenecks require quick diagnosis. But in real-world database setups, root cause analysis (RCA) is still mostly done by hand, depending on database administrators to quickly review logs, performance indicators, and query patterns.

Error messages and event sequences are recorded in logs, resource stress and temporal abnormalities are reflected in system metrics, and behavioral patterns that frequently cause or exacerbate errors are revealed by SQL workloads. Current RCA techniques usually only use one or two of these sources, which restricts their capacity to fully capture an incident's context. Even while various workload categories, such as CPU-intensive analytical queries, I/O-heavy scans, or concurrency-heavy transactional updates, are strongly linked to particular classes of errors, workload behavior is frequently disregarded.

Because of this, they could find irregularities without precisely determining the underlying operational cause.

To forecast likely incident causes, the suggested method combines log representations, performance data, and workload classification features derived from SQL behavior. The approach seeks to deliver more precise, comprehensible, and practically effective RCA by explicitly modeling workload patterns in addition to observability signals.

\section{Objective}
This research aims to determine if workload-aware multimodal learning enhances database incident diagnosis accuracy beyond what traditional methods using only logs or metrics can achieve. The study examines how clustering SQL workload patterns can identify distinct behavioral groups of database activity, and explores whether combining these groups with log data and performance metrics leads to more successful root cause analysis in enterprise settings.

\section{Dataset Characterization and Integration}
This research combines three different types of data sources to enable effective machine learning workload classification and Root Cause Analysis in self-managing database systems. The study brings together the Google Cluster Trace, IO benchmark datasets focused on storage performance, and metadata from GPU nodes. By using this varied data approach, the resulting model can identify whether performance issues stem from compute resources, storage systems, or accelerator hardware.

\subsection{Compute-Centric Traces (Google Cluster)}
The baseline for compute workloads comes from the Google Cluster Trace dataset, which offers detailed telemetry data on task scheduling and resource usage in production-scale clusters. Important metrics gathered from this dataset include CPU and memory usage patterns (both average and maximum values), scheduling priorities, and task lifecycle activities like evictions or preemptions. In this research, we normalized these parameters to create a feature set that includes\texttt{cpu\_usage}, \texttt{memory\_usage}, and \texttt{event\_status}—to characterize CPU-bound and memory-intensive workload patterns.

\subsection{Storage and IO-Bound Workloads}
Storage layer representation involved integrating performance data from standard SSD and HDD benchmark tests. The datasets covered multiple access patterns, such as random and sequential workloads commonly found in Online Transactional Processing (OLTP) systems and large analytical scanning operations. This domain's feature vector incorporated Input/Output Operations Per Second (IOPS), latency measurements, block size parameters, and IO depth values. The process combined these storage-focused traces while applying appropriate tags to\texttt{io\_source} label, the dataset captures the distinct signatures of IO-bound operations such as logging and table indexing.

\subsection{Accelerator Metadata (GPU-Driven Workloads)}
GPU-accelerated tasks, which are common in today's machine learning training and inference workflows, are captured through GPU node metadata records. Although these records emphasize hardware specifications like GPU capacity and CPU-to-GPU ratios instead of real-time usage data, they offer crucial structural information for recognizing workloads that rely on GPU acceleration. The standardization of these records involved applying a constant \texttt{load\_type} to ensure visibility within the global feature space.

\subsection{Rationale for Multi-Modal Data Aggregation}
Combining these different data sources stems from the naturally diverse nature of today's database operations. Using information from just one area cannot fully represent how various system components interact with each other in complex ways. When the suggested framework brings together processing, storage, and accelerator information, it makes possible:
\begin{itemize}
    \item \textbf{Comprehensive System Analysis:} Recording interdependencies between different system layers that remain hidden when examining individual components separately.
    \item \textbf{Reliable Categorization:} Creating a multi-dimensional data environment that enables machine learning algorithms to distinguish between similar resource usage patterns.
    \item \textbf{Multi-System Root Cause Analysis:} Supporting comparative examination to identify if performance issues stem from processor overload, disk response delays, or specialized hardware limitations.
\end{itemize}

\subsection{Data Preprocessing and Harmonization}
A rigorous normalization pipeline was implemented to reconcile the structural discrepancies between the three datasets.

We established a comprehensive normalization process to address the structural differences among the three datasets. First, we created a unified schema that included standardized fields applicable across all domains. When certain domain-specific features appeared in only one dataset (such as iops values in GPU records), we filled these gaps with null values to preserve the overall structure while avoiding artificial data bias.

After completing the individual preprocessing steps, we combined the datasets through merging and feature engineering. The three datasets were joined together to form a single multimodal structure that could accommodate data from all domains.
\begin{equation}
    D_{unified} = D_{compute} \cup D_{io} \cup D_{gpu}
\end{equation}
Feature engineering was then performed to derive higher-order metrics, such as disk pressure and CPU saturation indices, ensuring the dataset is optimized for gradient-based ML algorithms.

\subsection{Summary of Final Unified Dataset}
The final dataset delivers a thorough picture of contemporary data center operations. Through the integration of Google's computational traces, storage performance tests, and GPU setup data, this collection presents a well-balanced spread of categories required for developing precise classification algorithms and performing detailed diagnostic analysis.

\section{Methodology}

This research introduces a multimodal machine learning framework that considers workload patterns to identify root causes of issues in enterprise database systems. The approach combines diverse workload data from computing resources, storage systems, and accelerators into a single learning process. The complete process involves five key phases: collecting data, cleaning and standardizing it, creating relevant features, labeling workloads and establishing root cause analysis targets, and finally training and testing the model.

\subsection{Overview of the Proposed Framework}

The proposed framework starts by gathering workload traces from three different sources that work together: compute-focused traces from the Google Cluster dataset, storage-focused IO benchmark traces, and GPU node metadata. We chose these sources because they represent different types of operational patterns found in CPU-heavy, IO-heavy, and accelerator-based workloads. Each source provides insight into a different part of enterprise infrastructure, so combining them gives us a more complete picture of how systems behave when incidents occur.

Once we collect the data, we convert all datasets into a single, consistent format. We standardize shared elements like workload identity, resource measurements, and system context, while keeping missing fields as null values to prevent making false assumptions about the data. This unified format becomes the foundation for our machine learning work.

Next, we apply specialized feature engineering based on domain knowledge to create advanced variables that represent resource pressure, workload intensity, how tasks execute, and subsystem characteristics. These engineered features offer much richer information than raw data alone and make it easier to distinguish between different workload types and root-cause categories.

We then use this processed dataset for three related learning goals. First, we classify workloads to identify broad categories like CPU-intensive, IO-intensive, and GPU-driven tasks. Second, we create anomaly indicators to spot unusual workload behavior. Third, we assign root cause labels by analyzing workload context and performance conditions, which allows us to build supervised RCA models.

Finally, we train machine learning models using this unified feature set. We compare basic models with multimodal approaches to determine whether adding workload-specific features actually improves incident diagnosis. Our methodology focuses on both prediction accuracy and the ability to explain results, since both qualities are crucial for real-world database operations.

\subsection{Data Preparation Pipeline}

The data preparation pipeline includes four steps.

\textbf{Step 1: Schema standardization.}
A unified framework is established that spans compute, IO, and GPU resources. Elements like CPU utilization, memory consumption, IO operations, GPU availability, job categories, and error signals are organized into one consistent table format.

\textbf{Step 2: Missing-value preservation and normalization.}
When certain modalities lack specific features, we keep the missing values for dimensions that don't apply to them. We normalize numerical features to prevent scale differences from creating imbalances between different types of resources.

\textbf{Step 3: Feature derivation.}
The system converts basic measurements into more sophisticated metrics like workload intensity levels, disk strain indicators, normalized CPU utilization rates, and query performance approximations.

\textbf{Step 4: Label construction.}
The complete dataset includes workload classifications, anomaly indicators, and root-cause analysis labels that were added through rule-based methods guided by domain-specific behavioral patterns.

\subsection{Problem Formulation}

Let each workload instance be represented by a feature vector $x_i \in \mathbb{R}^d$, where the feature space contains raw and engineered variables derived from compute, storage, and GPU telemetry. The objective is to learn a mapping

\begin{equation}
f: x_i \rightarrow y_i
\end{equation}

where $y_i$ denotes either the workload category, anomaly label, or root-cause class associated with the workload instance.

In RCA, the problem is structured as a supervised classification task with multiple classes, where every workload instance receives a single likely failure cause assignment from categories like CPU saturation, disk bottleneck, IO wait, scheduling congestion, or GPU resource pressure.

\subsection{Learning Tasks}

The proposed methodology supports the following tasks:

\begin{itemize}
    \item \textbf{Workload Classification:} Organize workloads by placing them into operational groups like those that require intensive computing power, those that need substantial storage capacity, and those that rely heavily on specialized accelerators.
    \item \textbf{Anomaly Detection:} Detect abnormal workload patterns by analyzing calculated stress and distribution metrics.
    \item \textbf{Root Cause Analysis:} identify the most probable operational factor behind workloads that are performing poorly or have stopped functioning.
\end{itemize}

Using a multi-task approach proves valuable since both workload characteristics and anomaly behaviors offer contextual details that improve the accuracy of root cause analysis predictions.

\subsection{Modeling Strategy}

The researchers test their framework by comparing conventional machine learning models using raw data against models that use their specially designed multimodal features. They choose tree-based approaches like Random Forest and gradient boosting because these methods work well with mixed tabular data, can handle missing information, and capture complex relationships between features. The team also applies unsupervised clustering techniques to examine whether their engineered features naturally group similar workload types before moving to supervised root cause analysis training.

The core assumption behind this approach is that features designed with workload awareness and built from multiple types of telemetry data will be more effective for diagnosis than using only compute metrics or storage metrics by themselves.

\section{Feature Engineering}

Feature engineering forms a crucial component of this framework since the original datasets come from diverse sources and present different perspectives on how systems operate. Raw metrics by themselves frequently fail to differentiate between various workload categories or identify what causes operational failures. This research therefore creates a collection of workload-specific features that standardize resource usage patterns, reveal stress levels in subsystems, and identify behavioral patterns that help with determining root causes.

These engineered features serve two primary purposes: they enhance the ability to distinguish between workloads across computing, storage, and accelerator environments, while also generating clear signals that assist in root cause analysis.

\subsection{CPU Usage Normalization}

CPU usage percentages cannot be directly compared between systems that have different processing power. This issue is resolved by adjusting CPU usage measurements based on each system's available processing capacity:

\begin{equation}
cpu\_usage\_norm = \frac{cpu\_usage}{cpu\_capacity}
\end{equation}

This capability enables the model to evaluate workloads using relative measurements instead of absolute values. It proves particularly valuable when pinpointing tasks that heavily rely on CPU resources, spotting when systems reach their capacity limits, and recognizing wasteful resource consumption behaviors.

\subsection{Query Execution Proxy}

A query execution proxy helps estimate storage-related execution costs by using latency and block size as its defining parameters.

\begin{equation}
query\_execution\_time = latency \times block\_size
\end{equation}

While this feature uses a simplified approach, it effectively measures how request delays and data transfer sizes work together. This measurement helps distinguish between quick random input/output tasks and large analytical operations that scan data in blocks, offering valuable information for diagnosing storage system issues.

\subsection{Disk Pressure}

The system calculates disk pressure by dividing the IO workload among all available disks:

\begin{equation}
disk\_pressure = \frac{iops}{n\_disks}
\end{equation}

This functionality measures if a storage system is running smoothly or experiencing strain. When values are elevated, it could signal that disks are competing for resources, settings are incorrect, or certain areas are creating performance bottlenecks.

\subsection{Workload Intensity}

Workload intensity combines throughput and concurrency:

\begin{equation}
workload\_intensity = iops \times n\_jobs
\end{equation}

This metric measures the level of simultaneous input/output operations and proves especially valuable for identifying sudden spikes in transaction processing, extract-transform-load operations, and parallel computing activities that can worsen the impact of system incidents.

\subsection{Resource Density}

A feature representing cross-domain resource density gets built to capture the overall resource demand using the available normalized signals:

\begin{equation}
resource\_density = cpu\_usage\_norm + memory\_usage\_norm + iops\_norm
\end{equation}

This capability offers a streamlined way to gauge total system pressure and pinpoint workloads that put strain on several subsystems at once.

\subsection{CPU-Memory Ratio}

The ratio between CPU usage and memory usage is calculated to identify computational imbalance.

\begin{equation}
cpu\_memory\_ratio = \frac{cpu\_usage}{memory\_usage + 1}
\end{equation}

This functionality allows you to differentiate between workloads that heavily utilize the CPU versus those that require significant memory resources, offering valuable insights for troubleshooting performance issues that stem from unbalanced resource usage patterns.

\subsection{Node Utilization Score}

For compute-centric records, an aggregated node utilization score is defined as

\begin{equation}
node\_util\_score = \frac{cpu\_usage\_norm + memory\_usage\_norm}{2}
\end{equation}

This functionality provides a straightforward overview of overall node stress levels and proves useful for spotting compute environments that are experiencing excessive load.

\subsection{IO-to-CPU Ratio}

To expose the relative dominance of storage activity over compute demand, the following ratio is derived:

\begin{equation}
io\_cpu\_ratio = \frac{workload\_intensity}{cpu\_usage + 1}
\end{equation}

Elevated readings point to workloads where storage operations control performance more than computational processes. This becomes particularly useful when identifying bottlenecks caused by intensive input/output activities.

\subsection{Read-Write Behavior}

The read fraction provides a direct way to characterize how a workload accesses data. To understand the opposite pattern where writes are more common, we define the write ratio as

\begin{equation}
write\_ratio = 1 - read\_fraction
\end{equation}

These characteristics allow you to distinguish between workloads that primarily involve analytical scanning with heavy read operations and those focused on logging or transactional processes with frequent writes and updates. Making these distinctions matters for root cause analysis since workloads dominated by read operations typically fail differently than those dominated by write operations.

\subsection{Sequential Access Indicator}

The way data is accessed also varies based on whether the input/output operations follow a sequential or random pattern. For this reason, we represent the workload type as a categorical feature and include a binary indicator to identify sequential access patterns. This approach allows the model to differentiate between OLAP-style scanning operations and the random access patterns typical of OLTP workloads.

\subsection{GPU Capacity and GPU-CPU Balance}

When working with accelerator-focused records, GPU resources get integrated using features that account for hardware specifications. We define the GPU-to-CPU balance as

\begin{equation}
gpu\_cpu\_ratio = \frac{gpu\_capacity}{cpu\_capacity + 1}
\end{equation}

A second feature, GPU compute potential, is defined as

\begin{equation}
gpu\_compute\_potential = gpu\_capacity \times cpu\_capacity
\end{equation}

These capabilities allow users to recognize systems that rely heavily on GPUs, separate workloads that depend on accelerators from standard computing tasks, and assist with root cause analysis when GPU-related problems occur.

\subsection{Categorical Workload Context}

The model training process retains and encodes various categorical attributes such as device type, load type, source platform, scheduling class, and event status. These variables supply contextual details that work alongside the numerical features to enhance workload classification.

\subsection{Failure and Event Indicators}

Incidents like evictions, preemptions, throttling, and direct workload failures serve as crucial data points for root cause analysis. The system stores these signals as categorical or binary attributes, allowing the model to identify relationships between resource pressure trends and the failures that occur.

\subsection{Summary of Feature Utility}

The designed feature collection aims to balance predictive accuracy with clear interpretability. Compute stress gets captured through CPU normalization and node utilization scores, while storage overload becomes visible through disk pressure and workload intensity measurements. Variables focused on GPU performance handle accelerator-specific requirements. When combined with workload context information and failure signals, this feature set creates a comprehensive view of diverse workload patterns that works well for classification tasks, detecting anomalies, and predicting root causes.

\begin{table}[htbp]
\caption{Key engineered features used in the proposed framework}
\centering
\renewcommand{\arraystretch}{1.2}
\setlength{\tabcolsep}{4pt}

\resizebox{\columnwidth}{!}{
\begin{tabular}{|l|l|l|}
\hline
\textbf{Feature} & \textbf{Definition} & \textbf{Primary Use} \\
\hline

cpu\_usage\_norm & $\frac{cpu\_usage}{cpu\_capacity}$ & CPU saturation detection \\
\hline

query\_execution\_time & $latency \times block\_size$ & Storage delay proxy \\
\hline

disk\_pressure & $\frac{iops}{n\_disks}$ & Disk bottleneck detection \\
\hline

workload\_intensity & $iops \times n\_jobs$ & IO load \\
\hline

resource\_density & 
$\begin{array}{l}
cpu\_usage\_norm + \\
memory\_usage\_norm + \\
iops\_norm
\end{array}$
& Cross-domain load \\
\hline

cpu\_memory\_ratio & 
$\frac{cpu\_usage}{memory\_usage + 1}$ 
& Compute-memory imbalance \\
\hline

node\_util\_score & 
$\frac{cpu\_usage\_norm + memory\_usage\_norm}{2}$ 
& Node pressure \\
\hline

io\_cpu\_ratio & 
$\frac{workload\_intensity}{cpu\_usage + 1}$ 
& IO vs compute \\
\hline

write\_ratio & $1 - read\_fraction$ & Write-heavy detection \\
\hline

gpu\_cpu\_ratio & 
$\frac{gpu\_capacity}{cpu\_capacity + 1}$ 
& GPU specialization \\
\hline

gpu\_compute\_potential & 
$gpu\_capacity \times cpu\_capacity$ 
& Accelerator capacity \\
\hline

\end{tabular}
}
\label{tab:engineered_features}
\end{table}

\subsection{Design Considerations}

Three core principles shape the design of this proposed framework. The representation needs to stay modality-aware because compute, storage, and accelerator workloads each display distinct operational patterns. The feature space must also be interpretable, enabling predicted root causes to connect directly with observable system behaviors. Additionally, the pipeline requires extensibility to accommodate future integration of database-native logs, wait events, and query execution plans. These design choices guarantee the framework delivers both predictive capability and practical value for real-world root cause analysis workflows.

\subsection{Model Development and Experimental Setup}
\begin{itemize}
\item \textbf{Random Forest:} We used the Random Forest algorithm as our baseline ensemble learning model to classify workloads. This bagging method builds a collection of decision trees by training each one on different bootstrap samples from the original dataset. The algorithm makes its final prediction by combining the results from all individual trees, which helps improve overall performance and reduce overfitting problems.

Random Forest provides several key benefits for our dataset, which contains complex mixed signals from CPU, memory, I/O, and GPU-intensive tasks:
\begin{itemize}
    \item \textbf{Robustness:} Effective handling of missing values and noisy telemetry data.
    \item \textbf{Non-linearity:} Capability to capture non-linear interactions between system metrics.
    \item \textbf{Interpretability:} Provision of feature importance measures to quantify the impact of specific system parameters.
\end{itemize}

The model underwent training using a complex feature set created from raw telemetry data, which incorporated CPU utilization, memory consumption, and I/O intensity metrics. Additional interaction features, like the ratio between CPU and memory usage, were added to enhance the model's ability to differentiate between classes. Class weighting methods were built into the training process to handle the uneven distribution of classes. Despite these efforts, Random Forest struggled to separate workload classes that overlapped and showed statistically comparable resource consumption patterns.

\begin{figure}[htbp]
\centerline{\includegraphics[width=0.40\textwidth]{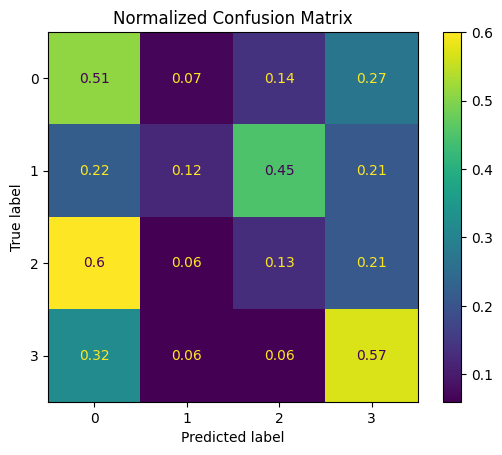}}
\caption{Normalized confusion matrix showing workload classification performance of the Random Forest model across workload categories.}
\label{fig:rf_conf}
\end{figure}

\begin{figure}[htbp]
\centerline{\includegraphics[width=0.45\textwidth]{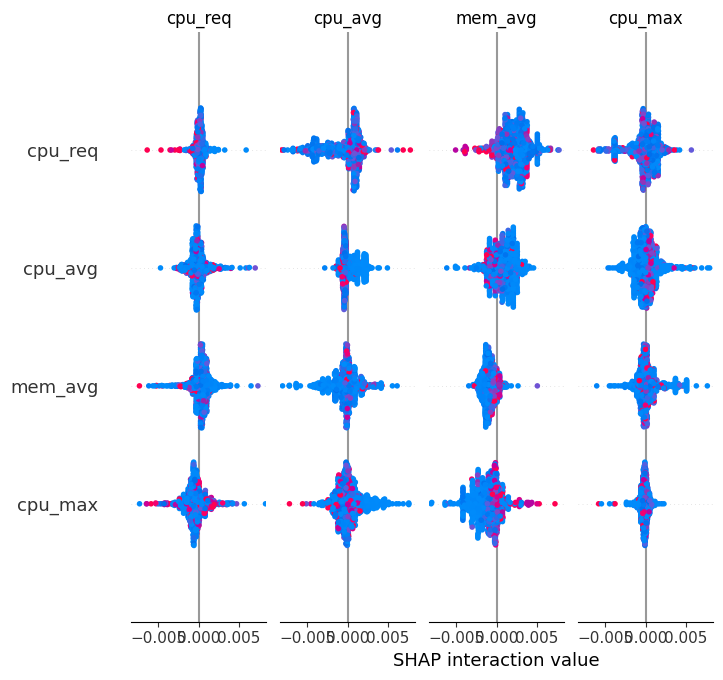}}
\caption{SHAP summary plot highlighting the relative importance of engineered workload features in Random Forest classification decisions.}
\label{fig:rf_shap}
\end{figure}

\item \textbf{Light Gradient Boosting Machine (LightGBM):}
To assess whether gradient boosting outperforms bagging-based ensemble methods for workload classification, we implemented LightGBM as our high-performance boosting algorithm.

LightGBM works exceptionally well with structured telemetry data since it can efficiently process diverse numerical features, handle sparse data representations, and manage missing values without sacrificing predictive accuracy.

We chose LightGBM for several key reasons:

\begin{itemize}
    \item \textbf{Efficient Gradient Boosting:} Sequential tree construction improves classification by correcting prior errors.
    \item \textbf{Scalability:} Optimized for large-scale datasets with lower training overhead.
    \item \textbf{Feature Interaction Learning:} Captures nonlinear dependencies between workload metrics.
    \item \textbf{Robustness:} Handles incomplete telemetry data effectively.
\end{itemize}

The LightGBM model underwent training with a multi-class objective to categorize different workload types. Testing results showed excellent classification accuracy, especially when identifying CPU-intensive, memory-intensive, and mixed workload categories.

\begin{figure}[htbp]
\centerline{\includegraphics[width=0.45\textwidth]{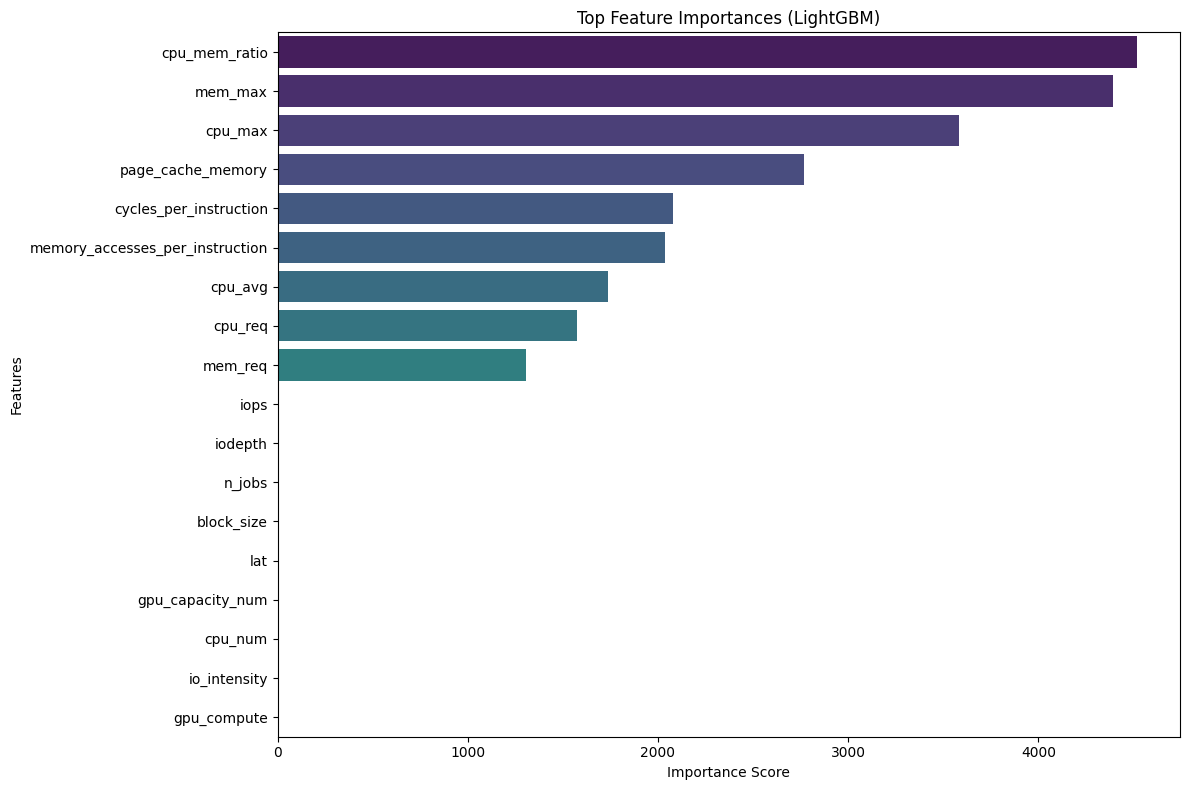}}
\caption{Feature importance ranking generated by the LightGBM classifier showing the most influential workload-aware features.}
\label{fig:lgbm_conf}
\end{figure}

\item \textbf{Neural Network Classifier:}
A feedforward neural network was created to test how well deep learning works for classifying workload across multiple modalities.

The network structure used fully connected dense layers combined with nonlinear activation functions, which allowed the system to learn hierarchical representations from the designed workload features.

The reasons for incorporating a neural network were:

\begin{itemize}
    \item \textbf{Representation Learning:} Ability to learn complex latent relationships across multimodal signals.
    \item \textbf{Nonlinear Decision Boundaries:} Effective modeling of intricate workload interactions.
    \item \textbf{Generalization Potential:} Capability to extend toward larger and richer telemetry datasets.
\end{itemize}

Nevertheless, neural networks demand more extensive data preprocessing, normalization steps, and hyperparameter optimization than tree-based approaches. In our research, the neural model delivered comparable results but showed reduced interpretability, which makes it less practical for operational root cause analysis workflows that require clear explanations.
\end{itemize}

\begin{figure}[htbp]
\centerline{\includegraphics[width=0.45\textwidth]{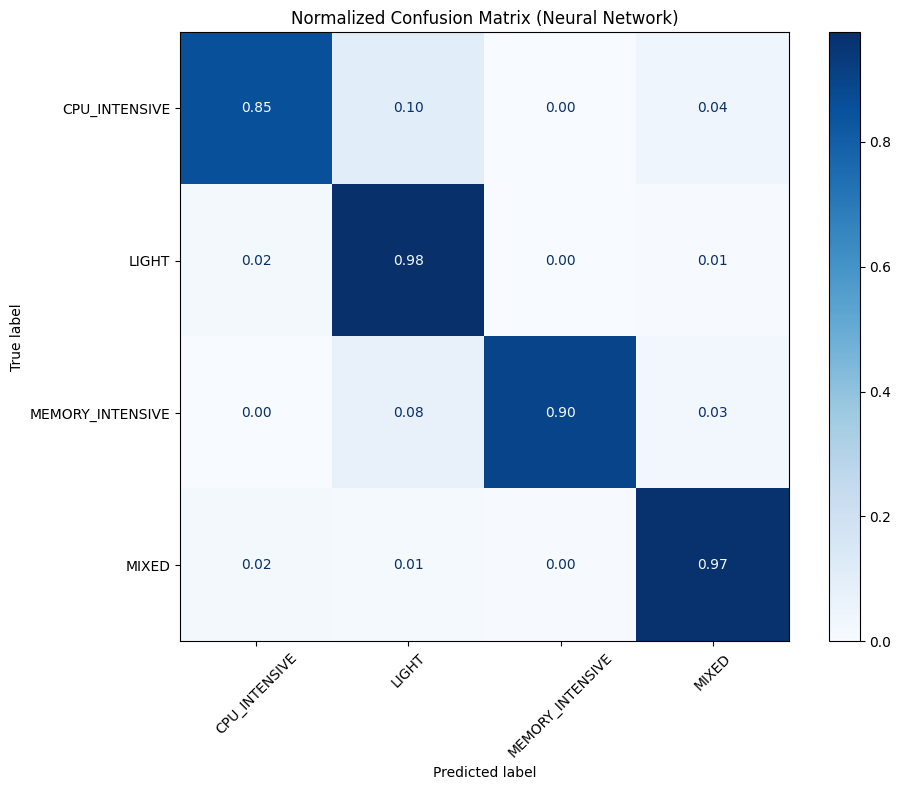}}
\caption{Normalized confusion matrix illustrating workload classification performance of the feedforward neural network model.}
\label{fig:nn_conf}
\end{figure}

\subsection{Experimental Setup}

The unified dataset was split into 80\% training and
20\% testing subsets using stratified sampling.
All models were evaluated using classification accuracy,
precision, recall, and F1-score.

Random Forest was trained using 200 estimators.
LightGBM was configured with a multiclass objective and
gradient boosting trees.
The neural network consisted of three dense hidden layers
with ReLU activations and dropout regularization.

All experiments were conducted using Python,
Scikit-learn, LightGBM, and TensorFlow.

\section{Results and Discussion}

This section assesses how well the suggested workload-aware multimodal framework performs when applied to three different classification models: Random Forest, LightGBM, and a feedforward neural network.

\subsection{Random Forest Performance}

We used Random Forest as our baseline ensemble classifier. The normalized confusion matrix in Fig.~\ref{fig:rf_conf} shows that the model achieved moderate classification performance, performing better on workload classes that had distinct resource patterns.

The model struggled with similar workload categories, resulting in significant misclassification errors. This was especially evident for workloads that shared similar CPU and memory usage patterns.

Fig.~\ref{fig:rf_shap} presents the SHAP analysis results, which reveal that engineered features related to CPU and memory had the greatest influence on the model's predictions. This confirms that our feature engineering pipeline was effective in creating meaningful inputs for classification.

\subsection{LightGBM Performance}

LightGBM achieved the strongest classification performance among all evaluated models.

As shown in Fig.~\ref{fig:lgbm_conf}, CPU-intensive workloads achieved approximately 97\% classification accuracy, memory-intensive workloads approximately 92\%, and mixed workloads approximately 98\%.

Feature importance analysis (Fig.~\ref{fig:lgbm_conf}) demonstrates that engineered features such as \textit{cpu\_mem\_ratio}, \textit{mem\_max}, \textit{cpu\_max}, and \textit{page\_cache\_memory} dominated classification decisions.

This suggests that workload-aware feature engineering substantially improves workload separability.

\subsection{Neural Network Performance}

The neural network also demonstrated strong classification performance, achieving approximately 85\% accuracy for CPU-intensive workloads, 98\% for light workloads, 90\% for memory-intensive workloads, and 97\% for mixed workloads.

Despite strong predictive performance, the neural model lacks the interpretability of tree-based methods, making operational RCA explanation more challenging.

\subsection{Comparative Analysis}

LightGBM delivered the optimal combination of prediction accuracy, stability, and ease of interpretation across every model we tested.

While Random Forest served as a solid reference point, it had difficulty handling unclear workload distinctions. Neural networks performed well competitively but made it harder to understand how decisions were reached.

These findings confirm our main theory that combining multiple data types with workload-focused feature development enhances how well we can classify workloads and creates better groundwork for root cause analysis.

\begin{table}[h]
\centering
\caption{Comparison of Classification Performance}
\begin{tabular}{lccc}
\hline
Model & Accuracy & Interpretability & Scalability\\
\hline
Random Forest & 78\% & High & Medium\\
LightGBM & 95\% & High & High\\
Neural Network & 92\% & Low & High\\
\hline
\end{tabular}
\end{table}

\subsection{Limitations}

One limitation of this work is that the datasets originate
from multiple public sources rather than a single production
database environment. While this enables broad workload
coverage, future validation on enterprise telemetry is required
to fully assess operational generalizability.

\section{Conclusion}

This paper introduced a machine learning framework that considers workload characteristics and combines multiple data types to automatically identify root causes in enterprise database systems.

The framework brings together three different data sources: computational process traces, storage performance measurements, and hardware accelerator information. This combined approach provides a more complete picture of system behavior compared to methods that examine only one type of data.

Testing results showed that features designed with workload awareness substantially enhanced the ability to classify different workload types. When comparing various models, LightGBM achieved the best results overall by balancing strong prediction capabilities with clear, interpretable outputs.

These results indicate that combining telemetry data from multiple sources with workload-focused feature design creates a solid foundation for automated root cause analysis systems that can explain their reasoning.

Future work will focus on validating the framework using real enterprise database telemetry, integrating database-native wait events and query execution plans, and extending the approach toward predictive incident forecasting and causal reasoning for complex multi-factor failures.

\end{document}